# A Novel Re-Targetable Application Development Platform for Healthcare Mobile Applications

**Chae Ho Cho[1], Fatemehsadat Tabei[2], Tra Nguyen Phan[3], Yeesock Kim[4] and Jo Woon Chong[5]**

[1, 2, 3, 5] Department of Electrical and Electronic Engineering, Texas Tech University, Lubbock, Texas 79409, USA.

[4] Department of Civil Engineering and Construction Management, California Baptist University, Riverside, CA 92504, USA.

[1]backjoe@gmail.com, [2]fatemehsadat.tabei@ttu.edu, [3]tra.phan@ttu.edu, [4]yekim@calbaptist.edu, [5]j.Chong@ttu.edu

**ABSTRACT**

The rapid enhancement of central power unit (CPU) performance enables the development of computationally-intensive healthcare mobile applications for smartphones and wearable devices. However, computationally-intensive mobile applications require significant application development time during the application porting procedure when the number of considering target devices' operating systems (OSs) is large. In this paper, we propose a novel re-targetable application development platform for healthcare mobile applications, which reduces application development time with maintaining the performance of the algorithm. Although the number of application's target OSs increases, the amount of time required for the code conversion step in the application porting procedure remains constant in the proposed re-targetable platform. Experimental results show that our proposed re-targetable platform gives reduced application development time compared to the conventional platform with maintaining the performance of the mobile application.

## 1. INTRODUCTION

Recently, mobile hardware manufacturers focus on developing wearable healthcare devices such as Fitbit [1] and Mio Band [2]. These wearable devices usually have simple biomedical signal processing software such as heart rate and respiratory rate detection software. However, the rapid enhancment of central power unit (CPU) performance enables the development of computationally-intensive mobile applications for smartphones and wearable devices. For example, computationally-intensive algorithms such as motion and noise artifacts (MNA) detection/reconstruction algorithms [3-8], arrhythmia detection algorithms [3-5, 9-10], and camera color correction algorithms [11] can currently run on smartphones. T2502A MediaTek (1.6 GHz 1 core). Moreover, Qualcomm's Snapdragon Wear 2100 (1.2 GHz quad-core) facilitate computationally-intensive algorithms to run on wearable devices.

As the healthcare algorithm gets computationally-intensive, the development of healthcare mobile applications takes longer time. Most of the development time is spent on designing algorithms, writing program code, porting (or building) program. Especially, porting procedure requires more time as the number of considering target devices' operating systems (OSs) increases. Hence, to decrease mobile application development time, efficient methods reducing the time required for this porting procedure is highly demanded. To the authors' knowledge, there has not been proposed an application development platform for healthcare mobile applications, which aims at efficiently reducing the time required for the porting procedure.

In this paper, we propose a re-targetable application development platform for healthcare mobile applications, which first converts written program code into one re-targetable program code, and then ports the converted one re-targetable program code into multiple target devices having different OSs. Hence, the proposed re-targetable platform can reduce the time required for the application porting procedure in mobile application development process. We evaluate our proposed re-targetable platform by comparing it to the conventional platform in terms of application development time and heart rate detection accuracy. The rest of this paper is organized as follows. Section 2 describes general application development process of mobile healthcare applications. In Section 3, our proposed re-targetable platform is explained comparing with conventional platforms. Section 4 evaluates the performance of our proposed re-targetable platform with comparing it to the conventional platform. Finally, Section 5 concludes this paper.

## 2. APPLICATION DEVELOPMENT PROCESS

The application development process of healthcare mobile applications generally consists of three procedure: data acquisition, algorithm development (designing algorithm and writing program code), and application porting procedures (see Fig. 1). Each of these procedures is described in Subsections 2.1, 2.2 and 2.3, respectively.

### 2.1 Data acquisition

Healthcare mobile applications for smartphones or wearable devices usually measure physiological information using sensors. In the data acquisition procedure, the raw signal is acquired to be analyzed to develop an effective healthcare algorithm. For instance, healthcare mobile applications detecting heart rhythm information measure raw physiological signals using sensors such as cameras, accelerometers, or microphones [7-11]. Here, when external sensors like electrocardiogram (ECG) sensors are used, measured physiological signals can be sent to smartphones or wearable devices through wireless communication techniques such as Wi-Fi, Bluetooth, and ZigBee. This acquired raw signal is analyzed to develop an effective algorithm for heart rhythm analysis.

### 2.2 Algorithm development

The algorithm development procedure consists of algorithm design and code writing steps. Healthcare algorithms are usually developed by high-level languages like MATLAB which enables developers to focus on what they want the algorithms to do since the high-level languages manage low-level programming details (e.g., memory management) in an automatic way. The high-level languages also have an advantage of facilitating algorithm development process since they mostly provide implemented functions in their libraries.

### 2.3 Application porting

The application porting procedure consists of converting, compiling and installing steps. The converting step first converts the source code developed by high-level language (see Subsection 2.2) appropriately into corresponding languages (iOS and Android) before porting it into target devices (iOS for iPhone and Android for Android phones). After this converting step, compiling and installing steps are performed appropriately according to target devices, which finalizes the application porting procedure.

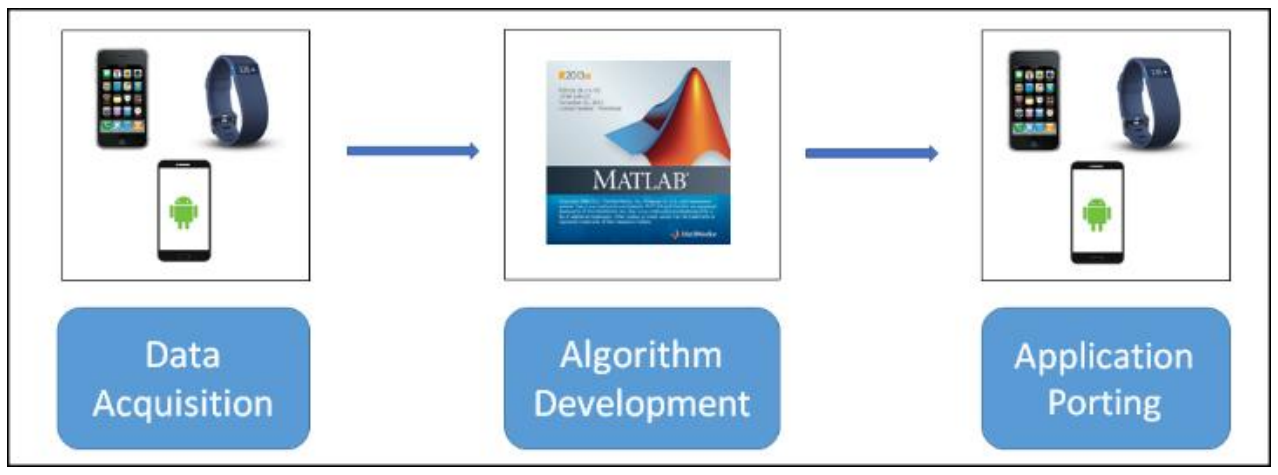

*Fig. 1. Application development process consisting of data acquisition (left), algorithm development (middle), and application porting (right) procedures.*

Table 1 shows a list of smartphones and wearable devices' OSs and their supporting main and sub-programming languages. For instance, healthcare mobile applications for Android are mainly developed using Java while the applications for iOS are developed using Objective-C. Hence, developers first need to convert the developed code in Subsection 2.2 to Java for Android and Objective-C for iOS. Consequently, converting step takes a longer time as the number of considering OSs increases.

The OSs usually supports C/C++ as their sub-programming language in addition to their main programming language as shown in Table 1. For instance, Java is the main programming language in Android OS but also C/C++ programming language is supported by Android OS. To enable interaction between Java language provided by Android software development kit (SDK) and C/C++ language provided by Android native development kit (NDK) [12], Android OS provides Java Native Interface (JNI) [13] as shown Fig. 2.

*Table 1: Main and sub-programming languages of smartphone and wearable device operating systems (OSs).*

<table>
<tr><th>Operating systems</th><th>Main Programming Language</th><th>Sub-Programming Language</th></tr>
<tr><td>Android</td><td>SDK (Java)<br>NDK (C/C++)</td><td rowspan="6">C/C++</td></tr>
<tr><td>iOS</td><td>Objective-C</td></tr>
<tr><td>Microsoft Windows</td><td>Multiple Languages</td></tr>
<tr><td>Real Time Operating System (RT OS)</td><td>C/C++</td></tr>
<tr><td>Linux</td><td>Multiple Languages</td></tr>
<tr><td>macOS</td><td>Multiple Languages</td></tr>
</table>

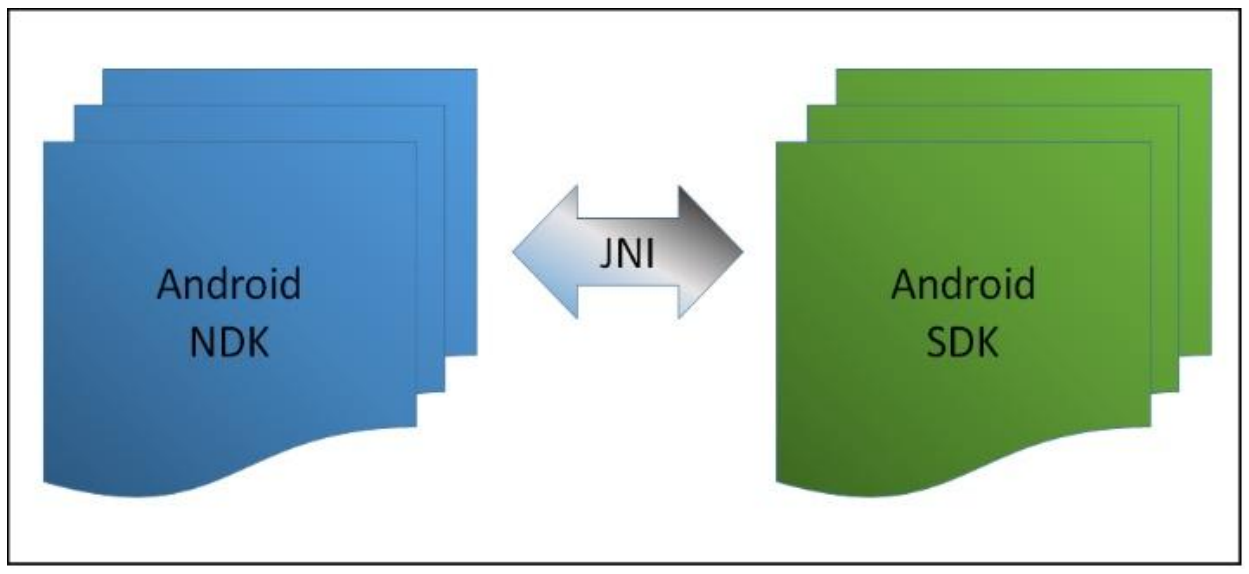

*Fig. 2. Java Native Interface (JNI) for interaction between SDK and NDK.*

# 3. PROPOSED RE-TARGETABLE PLATFORM

Our proposed re-targetable algorithm platform reduces algorithm development time by effectively decreasing developer's effort in the application porting procedure described in Section 2.3. Fig. 3 shows two major components to be ported in the application porting procedure: user interface (UI) part and algorithm part (or processing part). The algorithm part consists of algorithm component and algorithm interface. The algorithm component is mainly related to the program coded by main programming language while the algorithm interface is related to the functions coded by sub-programming language in Fig. 2.

The proposed re-targetable platform reduces the code conversion time of the algorithm component during the application porting procedure. Specifically, the proposed platform reduces this code conversion time by converting the sources code into the re-targetable code only once. Hence, the converted re-targetable code can be compiled and installed into any different OSs without any additional conversions.

Assuming that MATLAB is chosen as a high-level programming language and C/C++ is considered as a re-targetable programming language for developing healthcare applications, for instance, the code conversion procedure of the algorithm component in our proposed re-targetable platform can be further decomposed into main code, function code, and core mathematical code conversion steps (See Fig. 4.), which are described in Subsections 3.1, 3.2, and 3.3, respectively.

## 3.1 Main code

The main function in MATLAB needs to be converted into the re-targetable C/C++ code. Since they have different syntax and calling conventions, the healthcare algorithm

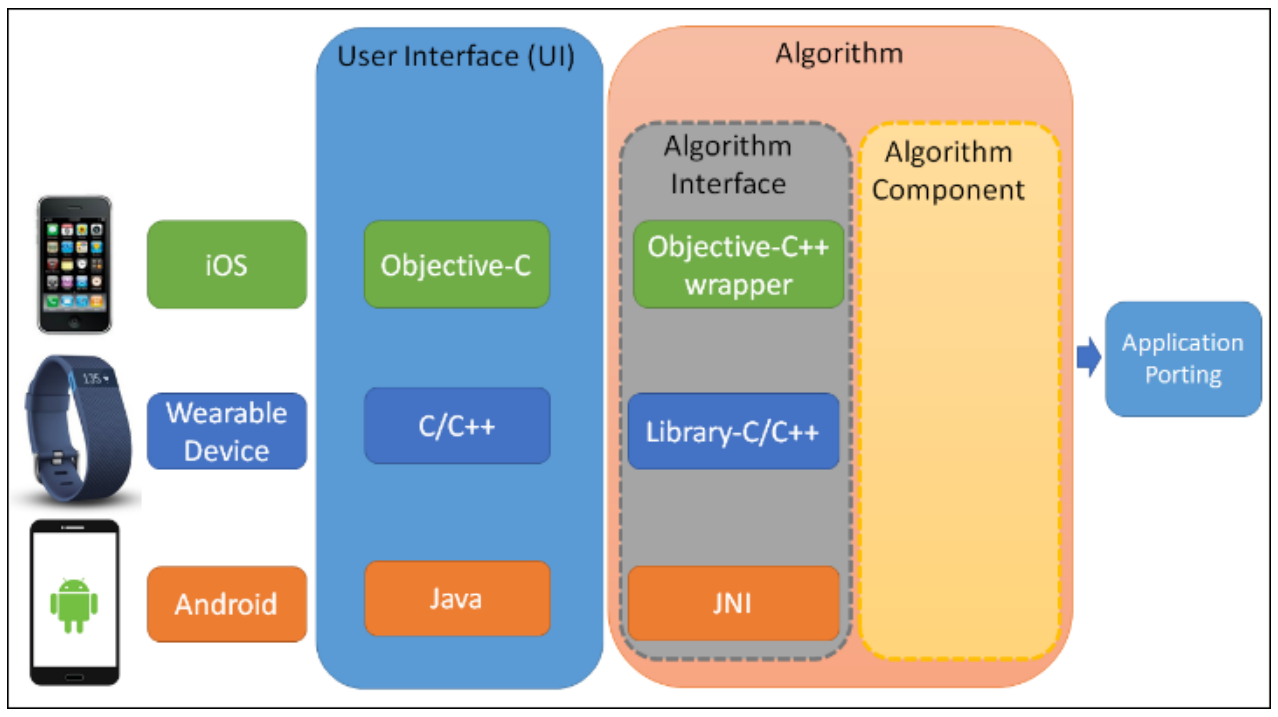

*Fig. 3. User interface and algorithm (or processing) components in the proposed re-targetable platform.*

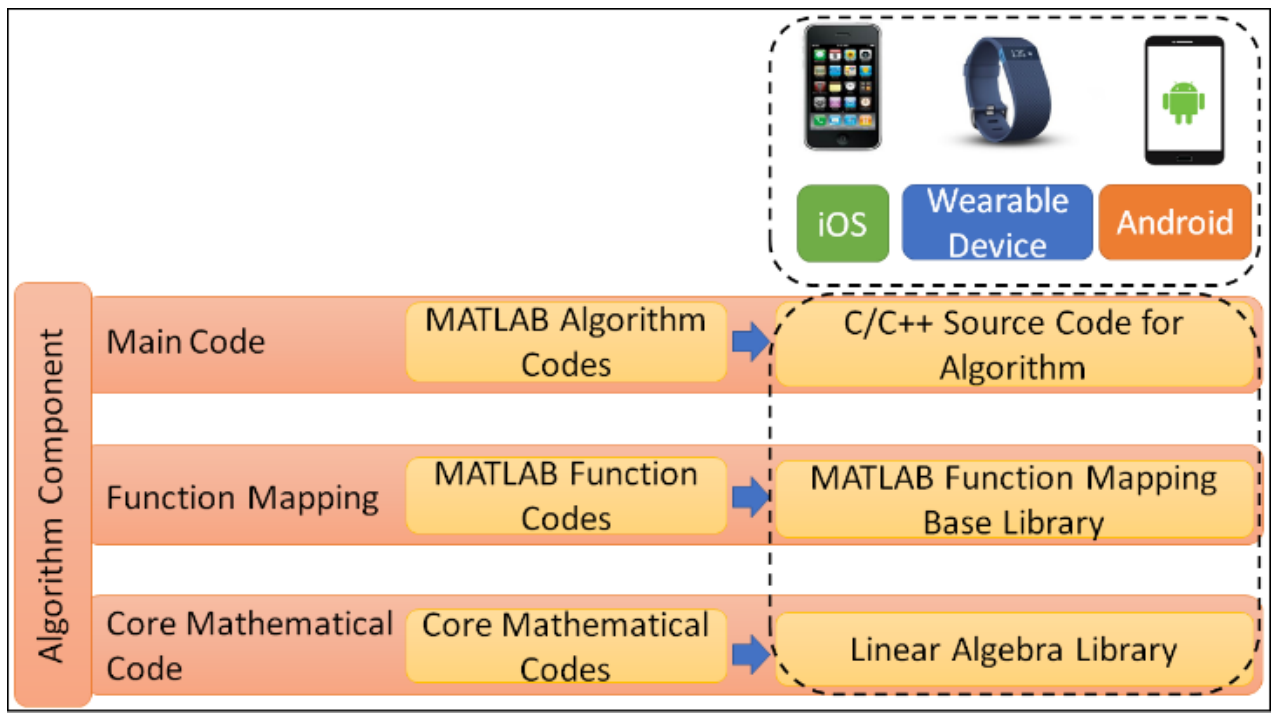

*Fig. 4. Three major code conversions steps in the algorithm layer: main code, function mapping, and core mathematical code conversions steps.*

developed by MATLAB code needs to be interpreted and converted appropriately into C/C++ code.

## 3.2 Function code

In addition to the main function, MATLAB provides numerous proven and implemented functions to developers. Since the function code may not implemented in C/C++ code, all the MATLAB functions used in healthcare algorithm need to be converted or mapped into re-targetable C/C++ code in an efficient way.

## 3.3 Core mathematical code

In the main and function code in MATLAB, core mathematical code such as linear algebra libraries handling vectors, matrices, cubes, floating point numbers and complex numbers may be used for healthcare applications. Hence, this core mathematical code may not exist or need to be converted in the re-targetable C/C++ language. Specifically, Armadillo [14], which provides a high-level syntax C++ library for C-friendly developers, provides numerous linear algebra-related classes. We adopt this Armadillo in our re-targetable platform.

# 4. RESULTS

We evaluate our proposed re-targetable platform by applying it to the application development process of electrocardiograph (ECG) heart rate detection application for Android phones and iPhones [15]. Here, ECG data is acquired by ECG sensors and sent to iPhone/Android using wireless technology. As performance metrics, we consider application development time and heart rate detection accuracy. Specifically, the heart rate is calculated from inversing average intervals between detected peaks of ECG signal and is compared to the heart rate calculated in a conventional way.

## 4.1 Application development time

We implement the ECG heart rate detection mobile application using our proposed re-targetable platform on Linux operating system. The build files related to the proposed re-targetable platform are shown in Fig. 5. Here, the re-targetable C/C++ code is developed once, and ported to both Android and iOS without repeating converting step of the application porting procedure (see Subsection 2.3). Android.mk build file helps installing the re-targetable C/C++ source code to the Android while Makefile is for installing the re-targetable code to iOS/Linux. The codes in Android.mk and Make build files are shown in Figs. 6 and 7, respectively.

Fig. 8 shows an example of converting MATLAB function code into re-targetable C/C++ code. MATLAB has a function named *filter* which filters an input signal using digital signal processing techniques. The re-targetable code in Fig. 8 was implemented by mapping Matlab *filter* function code into C/C++ code line-by-line on the C/C++ environment. This code conversion belongs to the function code conversion step of the proposed re-targetable platform

| | | | |
|---|---|---|---|
| Android.mk | Makefile | | main |
| Application.mk | algorithm | include | make |
| Config.mk | armadillo | clapack | lib |

*Fig. 5. Re-targetable platform directory architecture on Linux.*

```
LOCAL_PATH := $(call my-dir)
include $(CLEAR_VARS)
include $(DIR_PATH)/Config.mk

LOCAL_SRC_FILES += AFDetection.cpp
LOCAL_SRC_FILES += hermite_cubic.cpp
LOCAL_SRC_FILES += HighpassFilter.cpp
LOCAL_SRC_FILES += HRCalc.cpp
LOCAL_SRC_FILES += LowpassFilter.cpp
LOCAL_SRC_FILES += PsdWelch.cpp
LOCAL_SRC_FILES += WPIMatlabFunction.cpp
LOCAL_SRC_FILES += WPIMatrix.cpp
```

*Fig. 6. Makefile for Android (Android.mk).*

```
include ../make/build.inc

LOCAL_PRELINK_MODULE := false

OBJ1 += hermite_cubic.o
OBJ1 += HRCalc.o
OBJ1 += PsdWelch.o
OBJ1 += WPIMatlabFunction.o
OBJ1 += WPIMatrix.o
OBJ1 += filter.o
```

*Fig. 7. Makefile for iOS (Makefile).*

```
void MatlabFunc::filter(vectord B, vectord A, const vectord &X,
vectord &Y, vectord &Zi)
{
   double a0 = A[0];

   size_t input_size = X.size();
   size_t filter_order = (std::max)(A.size(), B.size());
   B.resize(filter_order, 0);
   A.resize(filter_order, 0);
   Zi.resize(filter_order, 0);
   Y.resize(input_size);

   const double *x = &X[0];
   const double *b = &B[0];
   const double *a = &A[0];
   double *y = &Y[0];
   double Y0 = X[0];
   double z[2000] = {0,0,0,0,0,0,0,};

   size_t order_N = filter_order - 1;

   for (size_t j = 0; j < order_N; j++ ){
         z[j] = Zi[j];
   }
   for (size_t i = 0; i < input_size; ++i)
   {
      y[i] = b[0] * x[i] + z[0];
      z[order_N] = 0;
      size_t order = 1;

      while (order < order_N+1)
      {
         z[order - 1] = b[order] * x[i] - a[order] * y[i] + z[order];
         ++order;
      }
   }
}
```

*Fig. 8. Re-targetable C/C++ ECG filter function code converted from MATLAB ECG filter function.*

described in Subsection 3.2.

Fig. 9 shows MATLAB code to obtain the heart rate from ECG using the Pan_Tomkins library [16-17] while Fig. 10 is its corresponding re-targetable C/C++ code converted using main code conversion step described in Subsection 3.1. Here, vector and matrix data type, and their operation, which are implemented by core mathematical code conversion described in Subsection 3.3, are also used.

### 4.2 Heart rate detection accuracy

Fig. 11 compares heart rate values of a subject derived from conventional and proposed re-targetable platforms. In the conventional platform, the heart rate detection application is developed following the general application development process described in Section 2. A subject is asked to do a normal activity from 0 to 2 seconds and to do running activity from 2 to 3.5 seconds. The heart rate values agree between the conventional and proposed platforms and heart rate increases when the subject is in the running activity.

```
function [iHR,tHR,peak] = EKGpeakDet(sig,fs)
L = length(sig);
[~,peak,~]=pan_tompkin(sig,fs,0);
if length(peak)>1
    RR = diff(peak);
else
    RR = 200;
end
iHR=60./RR*fs; % HR bpm
tHR=(peak(1:end-1)+RR/2)/fs;
```

*Fig. 9. MATLAB code for ECG peak detection*

```
void Algorithm::EKGpeakDet ( mat &sig, int fs, mat &iHR , vec
&peak, mat &tHR)
{
  vec RR ;
  uword L ;
  int delay = 0;
  mat qrs_amp_raw;
  vec qrs_i_raw;
  L = m2cpp::length(sig) ;
  pan_tompkin(sig, fs, qrs_amp_raw, peak, delay);
  if (m2cpp::length(peak)>1)
  {
    RR = diff(peak) ;
  }
  else
  {
    RR = 200 ;
  }
  iHR = 60.0*1.0/RR*fs ;
  tHR = (peak(m2cpp::span<uvec>(0, peak.n_rows-
2))+RR/2.0)/fs ;
}
```

*Fig. 10. Re-targetable C/C++ code converted from MATLAB ECG peak detection code in Fig. 9.*

## 5. CONCLUSION

In this paper, we have proposed a re-targetable platform for healthcare mobile application development, which reduces application development time with maintaining the performance of the application. Similarly to general mobile application process in conventional platform, the application development process in our proposed re-targetable platform follows three procedures: data acquisition, algorithm development, and application porting.

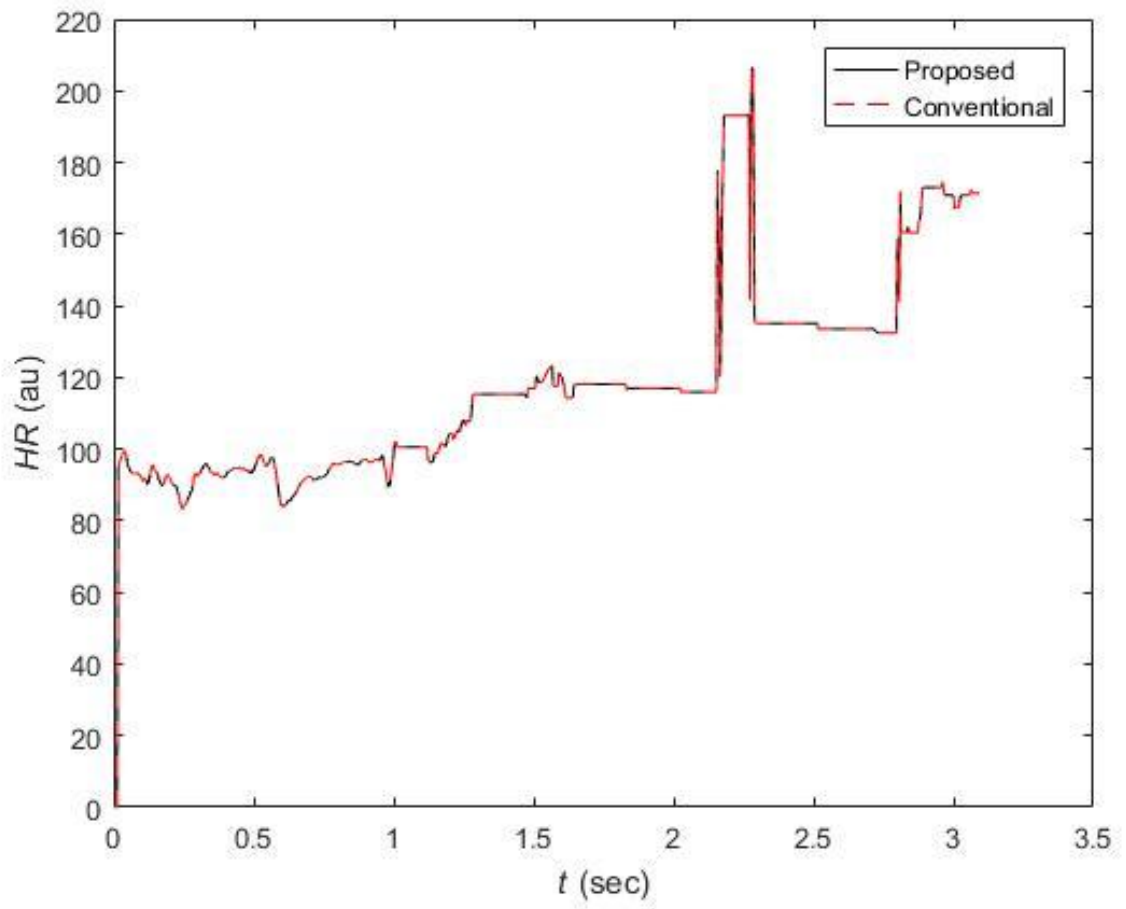

*Fig. 11. Heart rate calculated from proposed (black, solid) and conventional (red, dashed) platforms.*

However, the conventional platform requires developers to convert source code into multiple corresponding languages (e.g. Java for Android, Objective-C for iOS) in the application porting procedure while the proposed platform requires only one conversion from the source code to re-targetable one (C/C++). We evaluated our proposed re-targetable platform by applying it to developing ECG heart rate detection mobile applications for Android phones and iPhones. The results showed that the proposed re-targetable platform gives reduced development time compared to the conventional platform. Moreover, the heart rates of a subject detected from the proposed and conventional platforms were identical in both normal and running activity. We expect that the proposed re-targetable platform can be applied to other devices having different operating systems such as Windows, RT OS, Linux, and macOS.

## Acknowledgments

This work was supported by the National Institute of Health under Grant 1R15HL121761-01A1.